\documentclass[twocolumn, trackchanges]{aastex61}

\usepackage[fleqn]{amsmath}
\makeatletter
\newcommand*{\rom}[1]{\expandafter\@slowromancap\romannumeral #1@}
\makeatother

\newcommand{\vb}{\mathbf{b}}

\newcommand{\vbeta}{\boldsymbol{\beta}}
\newcommand{\vtheta}{\boldsymbol{\theta}}

\newcommand{\calGP}{\mathcal{GP}}

\newcommand{\calN}{\mathcal{N}}

\newcommand{\vc}{\mathbf{c}}

\newcommand{\calI}{\mathcal{I}}
\newcommand\scalemath[2]{\scalebox{#1}{\mbox{\ensuremath{\displaystyle #2}}}}

\shorttitle{NIR Mira PLRs in M33} 
\shortauthors{Yuan et al.}

\begin{document}
\title{Near-infrared Mira Period--Luminosity Relations in M33}

\author[0000-0001-9420-6525]{Wenlong Yuan}
\affiliation{George P.~and Cynthia W.~Mitchell Institute for Fundamental Physics \& Astronomy, Department of Physics \& Astronomy,\\Texas A\&M University, College Station, TX, USA}
\affiliation{Department of Physics \& Astronomy, Johns Hopkins University, Baltimore, MD, USA}
\author[0000-0002-1775-4859]{Lucas M.~Macri}
\affiliation{George P.~and Cynthia W.~Mitchell Institute for Fundamental Physics \& Astronomy, Department of Physics \& Astronomy,\\Texas A\&M University, College Station, TX, USA}
\author{Atefeh Javadi}
\affiliation{School of Astronomy, Institute for Research in Fundamental Sciences (IPM), P.O. Box 19395-5531, Tehran, Iran}
\author{Zhenfeng Lin}
\affiliation{Department of Statistics, Texas A\&M University, Texas, USA}
\author{Jianhua Z.~Huang}
\affiliation{Department of Statistics, Texas A\&M University, Texas, USA}
\email{lmacri@tamu.edu}

\begin{abstract} 
We analyze sparsely-sampled near-infrared ($JHK_s$) light curves of a sample of 1781 Mira variable candidates in M33, originally discovered using $I$-band time-series observations. We extend our single-band semi-parametric Gaussian process modeling of Mira light curves to a multi-band version and obtain improved period determinations. We use our previous results on near-infrared properties of candidate Miras in the LMC to classify the majority of the M33 sample into Oxygen- or Carbon-rich subsets. We derive Period-Luminosity relations for O-rich Miras and determine a distance modulus for M33 of $24.80\pm0.06$ mag.
\end{abstract} 

\section{Introduction}
Mira variables (Miras) belong to a class of long-period pulsators with large-amplitude cyclical luminosity variations~\citep{Kholopov1985} that also exhibit cycle-to-cycle and long-term magnitude changes~\citep{Mattei1997}. Miras can be further subdivided into Oxygen- or Carbon-rich (O- \& C-rich, respectively) based on their photospheric abundances and/or SEDs. O-rich Miras exhibit relatively tight Period--Luminosity relations (PLRs) \citep{Glass1981, Feast1989, Wood1999, Whitelock2008, Yuan2017b}, making them promising distance indicators for extragalactic systems \citep[e.g.][]{Whitelock2013, Whitelock2014, Menzies2015, Huang2018}.

A recent study~\citep{Yuan2017a} used decade-long $I$-band time-series photometry to discover 1847 Mira candidates in M33, the third largest spiral galaxy in the Local Group. Asymptotic giant branch variable stars in this system have been studied by several authors~\citep[e.g.,][]{Hartman2006,McQuinn2007,Cioni2008}, but none of them obtained a Mira-based distance. The distance modulus of this system has been previously determined by various means (other than Miras), of which we only highlight a few. Studies based on classical Cepheids have found $24.65\pm0.12$~mag~\citep{Macri2001a}, $24.53\pm0.11$~mag~\citep{Scowcroft2009} and $24.62\pm0.07$~mag~\citep{Gieren2013}. Detached eclipsing binaries were used by \citet{Bonanos2006} to obtain $24.92\pm0.12$~mag, while RR Lyrae yielded $24.67\pm0.08$~mag~\citep{Sarajedini2006}.

Given its relatively nearby distance, large sample of Miras, and available time-series photometry, M33 is well-suited for testing algorithms to analyze sparsely-sampled light curves and for characterizing the near-infrared (NIR) PLRs of these variables. In this work, we collect NIR and optical measurements of M33 Miras from various sources, analyze them using a novel technique, and derive a precise Mira-based distance to this galaxy. The rest of the paper is organized as follows. Section~\ref{sec_data} describes the data used in this study. In Section~\ref{sec_period} we introduce the method of period redetermination using multiband data and evaluate its accuracy on sparse Mira light curves. Section~\ref{sec_mag} describes the derivation of mean NIR magnitudes and the estimation of their errors. We present our results in Section~\ref{sec_result}.

\section{Data}\label{sec_data} 

\begin{figure}
\epsscale{1.0}
\plotone{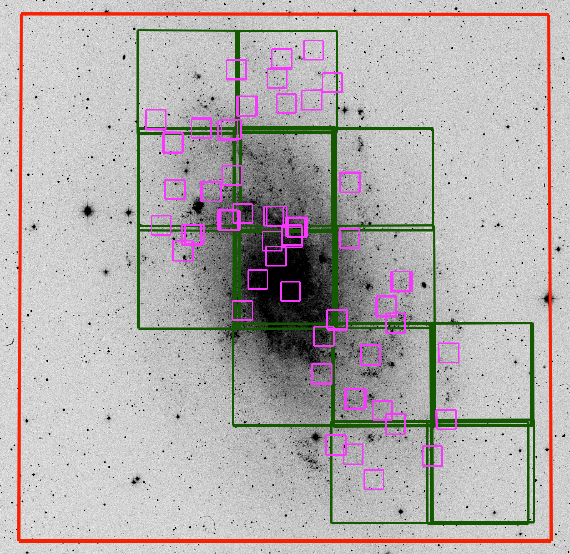}
\caption{Locations of UKIRT (red), KPNO (green), and Gemini (magenta) fields on a Digitized Sky Survey image of M33. The fields roughly cover 0.81, 0.37, and 0.05 square degrees, respectively. North is up and east is to the left. \label{fig_obs}}
\end{figure}

We based our study on NIR observations of M33 with: (1) the 3.8~m UK InfraRed Telescope~\citep[UKIRT, previously published by][]{Javadi2015}, (2) the 4-m Mayall telescope at Kitt Peak National Observatory (KPNO, published here for the first time) and (3) the 8-m Gemini North telescope (also published here for the first time). We further make use of previously-published $I$-band time-series photometry obtained by the DIRECT project and follow-up observations~\citep{Macri2001,Pellerin2011}, which were analyzed to search for Mira variables by~\citet{Yuan2017a}. The sky coverage of the aforementioned NIR surveys are shown in Figure~\ref{fig_obs}. We cross-matched stellar sources \citep[including Mira candidates from][]{Yuan2017a} among these datasets by updating all their astrometry to a single reference frame (defined by the UKIRT catalog). In the rest of the section, we describe the observations, data reduction, photometry, and astrometric calibration for each dataset.

\subsection{UKIRT Observations}

We used the photometrically and astrometrically calibrated data products of the UKIRT observations described in~\citet{Javadi2015}. These include multi-epoch measurements of the entire disk of M33 in $JHK_s$ from 2005 September to 2007 October. Aperture photometry was performed on the images, followed by photometric calibration using the 2MASS catalog. For the $\sim 2.6\times10^5$ sources with $\sigma_K < 0.2$~mag in the catalog, the median number of measurements per star are 2, 3, and 7 in $J$, $H$, and $K_s$, respectively. We refer the interested readers to~\citet{Javadi2015} and \citet{Hodgkin2009} for details of the observations, data reduction, photometry, and astrometric calibration.

\vfill\pagebreak\newpage

\subsection{KPNO Observations}

We used the FLAMINGOS NIR Imager~\citep{1998SPIE.3354..404E} on the KPNO 4-m telescope to observe 13 fields covering most of the disk of M33. FLAMINGOS had a 10\arcmin$\times$10\arcmin~field of view projected onto a 2K$\times$2K HgCdTe detector, which yields a pixel scale of $0\farcs316$/pixel. We obtained $JHK_s$ photometric measurements for each field on two consecutive nights. All the images were taken using a 3$\times$3 dither pattern with total exposure times of 10s or 30s. A summary of the observations is given in Table~\ref{tbl_obs}.

The raw images were processed using the {\tt MSCRED} package in {\tt IRAF}. We performed a two-pass source detection at a 2.5$\sigma$ threshold using {\tt DAOPHOT}~\citep{Stetson1987}, followed by PSF photometry using {\tt ALLSTAR}~\citep{Stetson1994}. Approximately 60000 stellar sources were found with $\sigma_K<0.2$~mag. We attempted to match these against the UKIRT catalog and found unacceptable results if we used a conventional WCS projection (i.e., CD matrix). Thus, we fit the distortion of the KPNO images using thin-plate spline models and achieved a typical WCS precision of $0\farcs17$. Using this nonparametrically calibrated WCS, we were able to match up sources in overlapping regions of KPNO fields as well as the overall KPNO star list with respect to the UKIRT catalog. We calibrated the KPNO photometry using the UKIRT catalog as reference, once again using thin-plate splines to account for zeropoint variations across the image plane. The calibration was based on $54-64$ stars spanning $0.3<J-K_s<2$. No significant color terms were found for these transformations (see Appendix for details).

\subsection{Gemini Observations}

We used the Gemini North NIR Imager~\citep[NIRI,][]{2003PASP..115.1388H} in f/6 mode to obtain $JHK_s$ observations of 46 small fields (2\arcmin$\times$2\arcmin) across the disk of M33. While these fields were originally selected to maximize the number of Cepheids from \citet{Macri2001a}, they also included many Mira candidates that were unknown at the time of the observations. Images were obtained on 12 different nights from 2002 September to 2006 January, with a few fields observed multiple times. The exposure times were 30.1s in all bands. Details of these observations are also given in Table~\ref{tbl_obs}.

The raw images were processed using the {\tt Gemini} package in {\tt IRAF}, while the subsequent steps were identical to those described for the KPNO observations. The Gemini images were substantially deeper than those from the 4-m telescopes, and we therefore used a slightly higher threshold (3.5$\sigma$) for source detection. We cross-matched the Gemini stellar sources to the UKIRT catalog to obtain the photometric calibration, using $160-215$ bright and isolated stars spanning $0.3<J-K_s<2$ to determine the color terms (see Appendix for details).

\subsection{DIRECT and Follow-up Observations}

We retrieved the $I$-band photometric measurements and the Mira candidate catalog from~\citet{Yuan2017a}. The $I$-band observations were collected by the DIRECT survey~\citep{Macri2001} and follow-up observations~\citep{Pellerin2011} with a combined baseline of nearly a decade. We re-calibrated the WCS coordinates of the sources using the UKIRT catalog as a reference in order to easily identify Mira candidates among different datasets. We refer the interested readers to~\citet[][and references therein]{Pellerin2011} for details of the observations, data reduction, and photometry of the $I$-band observations. The Mira catalog contains 1847 candidates, of which 1781 were found to have NIR measurements. The other 66 objects were excluded from this study.

\begin{deluxetable}{ccccccc}
\tabletypesize{\scriptsize}
\tablecaption{Observing Log\label{tbl_obs}}
\tablewidth{0pt}
\tablehead{
\colhead{Date$^a$} & \colhead{Site} & \colhead{Field} & \colhead{Band} & \colhead{R.A.} & \colhead{Dec.} & \colhead{Exp.} \\
\colhead{[day]} & \colhead{} & \colhead{} & \colhead{} & \multicolumn{2}{c}{[deg]} & \colhead{[sec]}}
\startdata
2536.959 & Gemini & G62 &$H$   & 23.36129 & 30.58773 & 30.1 \\
2536.966 & Gemini & G62 &$K_s$ & 23.36129 & 30.58774 & 30.1 \\
2536.980 & Gemini & G64 &$H$   & 23.44946 & 30.72921 & 30.1 \\
2536.986 & Gemini & G64 &$K_s$ & 23.44944 & 30.72921 & 30.1 \\
2537.003 & Gemini & G65 &$H$   & 23.48259 & 30.69734 & 30.1 \\
2537.010 & Gemini & G65 &$K_s$ & 23.48261 & 30.69732 & 30.1 \\
\enddata
\tablecomments{$a$: JD - 2,450,000. This table is available in its entirety in machine-readable form.}
\end{deluxetable}

\section{Period Search}\label{sec_period} 

Combining the $I$-band light curves with the NIR data, we redetermined the periods for the Mira candidates discovered by \citet{Yuan2017a}. We extended the semi-parametric Gaussian process model developed by \citet{He2016} to a multiband version. We evaluated the multiband model on simulated light curves with the same noise and sampling of the combined M33 data.

\subsection{Multiband Semi-Parametric\\Gaussian Process Model}

To describe the Mira light curves, which exhibit unpredictable cycle-to-cycle and long-term variations, \citet{He2016} developed a semi-parametric Gaussian process model which decomposes the $I$-band light curve into strictly periodic and data-driven components. The period is solved by optimizing the likelihood of the model fit. We refer interested readers to \citet{He2016} for a detailed description of the model. We extended the model to a multiband version to simultaneously fit sparsely-sampled $I$ and NIR light curves. Due to the limited amount of NIR measurements, it is desirable to minimize the number of free parameters in the model. Therefore, we fixed the amplitude ratios and phase lags of the periodic component among different bands, obtained from a prior study (see the end of this subsection for details).

For a set of multiband time-series data {$t_i, y_i, \sigma_i, \lambda_i$}, where $t_i$ is the time of the $i^{\rm th}$ observation, $y_i$ is the measurement, $\sigma_i$ is the measurement uncertainty, and $\lambda_i$ is the band, the multiband semi-parametric Gaussian process model can be expressed as
\begin{align*}
y_i | &\vbeta, g(t_i)  \sim \calN(g(t_i), \sigma_i^2), \\
& g(t)  = X\vbeta + h(t), \\
& \vbeta  \sim \calN(\vc,\sigma_c^2I_6),\\
& h(t)| \vtheta \sim  \calGP\Big(0,k_{\vtheta} (t,t') \cdot \calI(\lambda, \lambda')\Big), \\
& k_{\vtheta} (t,t') = \theta_1^2\exp \left(-\frac{(t-t')^2}{2\theta_2^2}\right), \\
& \calI(\lambda, \lambda') = \begin{cases} 1 \text{ if $\lambda=I$ and $\lambda'=I$} \\
1 \text{ if $\lambda \in JHK_s$ and $\lambda' \in JHK_s$} \\ 
0 \text{ for all other cases} \end{cases}.
\end{align*}
We replaced the periodic term $m + \vb_f(t)^T\vbeta$ in Eqn.~10 of ~\citet{He2016} with a linear multiband expression $X\vbeta$. $X$ is a design matrix that incorporates fixed amplitude ratios and phase lag relations among the $IJHK_s$ bands. Four representative rows are shown below to illustrate its formulation:
\[
\scalemath{0.9}{
  \left[ {\begin{array}{cccccc}
            \cos(\omega t_I) &  \sin(\omega t_I)  & 1 & 0 & 0 & 0\\
            \frac{A_J}{A_I} \cos(\omega t_J-\Delta\Phi_{IJ}) & \frac{A_J}{A_I} \cos(\omega t_J-\Delta\Phi_{IJ}) & 0 & 1 & 0 & 0\\
            \frac{A_H}{A_I} \cos(\omega t_H-\Delta\Phi_{IH}) & \frac{A_H}{A_I} \cos(\omega t_H-\Delta\Phi_{IH}) & 0 & 0 & 1 & 0\\
            \frac{A_K}{A_I} \cos(\omega t_K-\Delta\Phi_{IK}) & \frac{A_K}{A_I} \cos(\omega t_K-\Delta\Phi_{IK}) & 0 & 0 & 0 & 1\\
  \end{array} } \right],
}
\]
where $\omega\equiv 2\pi/P$, $P$ is the Mira period, $t_\lambda$ indicates the observation at time $t$ is through band $\lambda$ (one of $IJHK_s$), $A_\lambda$ is the light curve amplitude in the given band, and $\Delta\Phi_{I\lambda}$ is the phase lag between $I$ and $\lambda$. The number of rows of the design matrix is equal to the total number of measurements in all bands. Similar to \citet{He2016}, we used a Gaussian prior for $\vbeta$ with wide variance $ \vbeta \sim \calN(\vc,\sigma_c^2 I_6)$ where $\vc = (0, 0, \overline{m}_I, \overline{m}_J, \overline{m}_H, \overline{m}_K)$.

For the data-driven component, we assumed that the $JHK_s$ bands exhibit the same cycle-to-cycle and long-term variations, which are distinct from those of the $I$ band. We made this choice based on several facts: (1) our $I$ observations do not generally overlap in time with the NIR observations, so the data in the former do not drive the model for the latter and vice versa; (2) our study of Miras in the LMC \citep{Yuan2017b} shows that light curve variations in $JHK_s$ are very similar, while those in $I$ are usually much greater; (3) our $JHK_s$ data are extremely sparse and it is not feasible to solve for individual aperiodic variations. We further assumed that the hyperparameters $\theta_1$ and $\theta_2$ are the same across all bands under consideration, which should be valid as long as the strength and time scale of the covariance in the aperiodic process is the same for these wavelengths. We implemented these choices by multiplying the squared exponential kernel with a scalar function $\calI(\lambda, \lambda')$, which is set to a value of 1 if both observations are in $I$ or in any of $JHK_s$. Otherwise, $\calI(\lambda, \lambda') = 0$.

We note that this multiband model also works for objects with observations in fewer than four bands without any modifications. We adopted the same strategy as H16 to compute and optimize the likelihood. For a set of multiband light curves with unknown period, we compute the log-likelihood of the model fit for a dense grid of trial periods and adopt the resulting profile as the periodogram.

\begin{figure*}
\includegraphics[width=\textwidth]{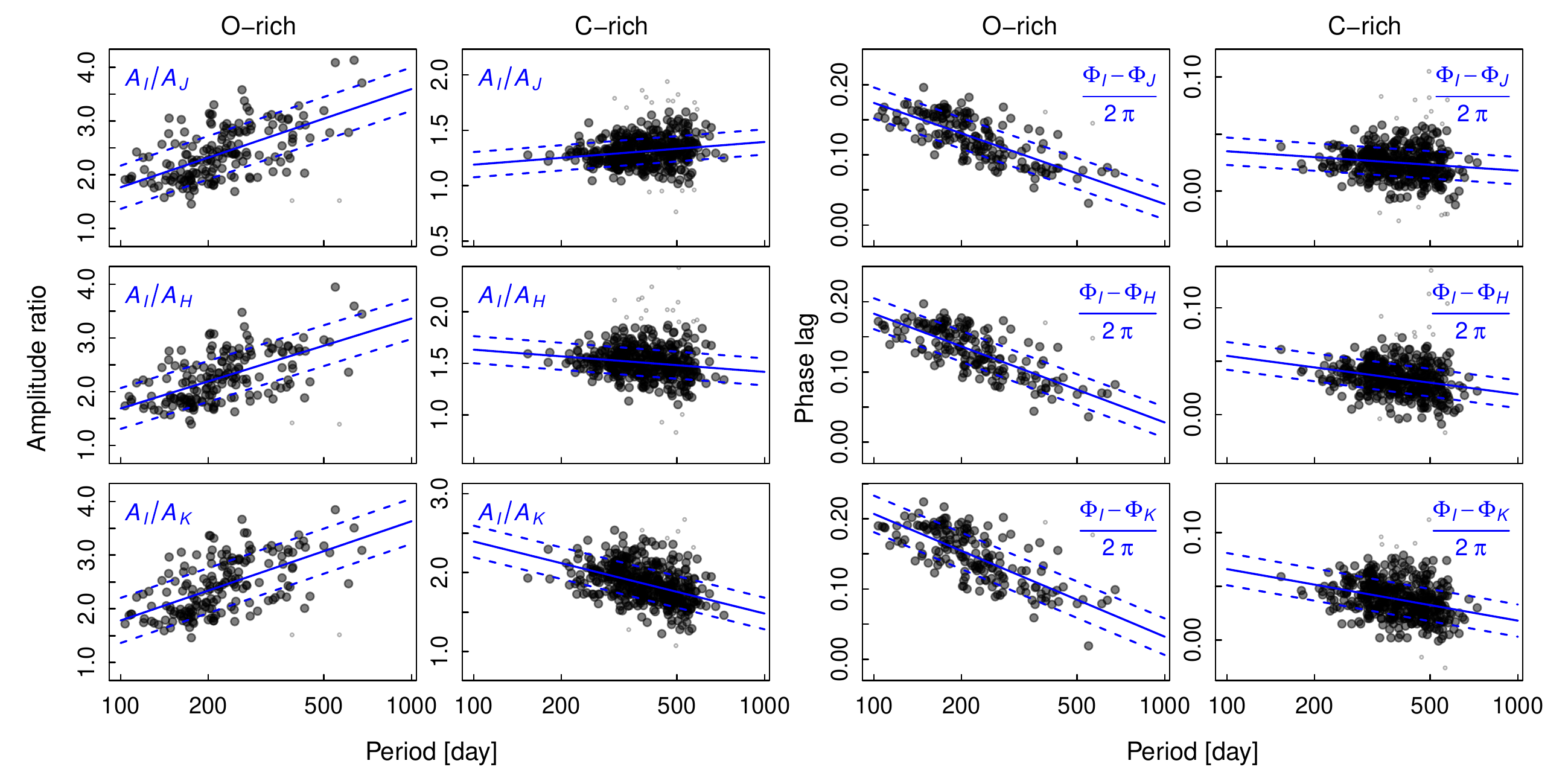}
\caption{$I$-to-$JHK_s$ amplitude ratios and phase lags derived from LMC Miras. The blue lines show the least-square fits and 1$\sigma$ uncertainties against $\log P$. Smaller points indicate outliers rejected by iterative 3$\sigma$ clipping. \label{fig_scale}}
\end{figure*}

To determine the fixed amplitude ratios and phase lag relations used in the multiband model, we made use of the OGLE-\rom{3} LMC Mira $I$-band light curves~\citep{Soszynski2009} and their $JHK_s$ template light curves derived by \citet{Yuan2017b}, who computed NIR template curves for individual LMC Miras based on 3 epochs of observations by \citet{Macri2015}. We fit first-order functions of $\log P$ to the $I$-to-$JHK_s$ amplitude ratios and phase lags. The results are shown in Figure~\ref{fig_scale} and Table~\ref{tbl_scale}. 

\begin{figure*}
\includegraphics[width=\textwidth]{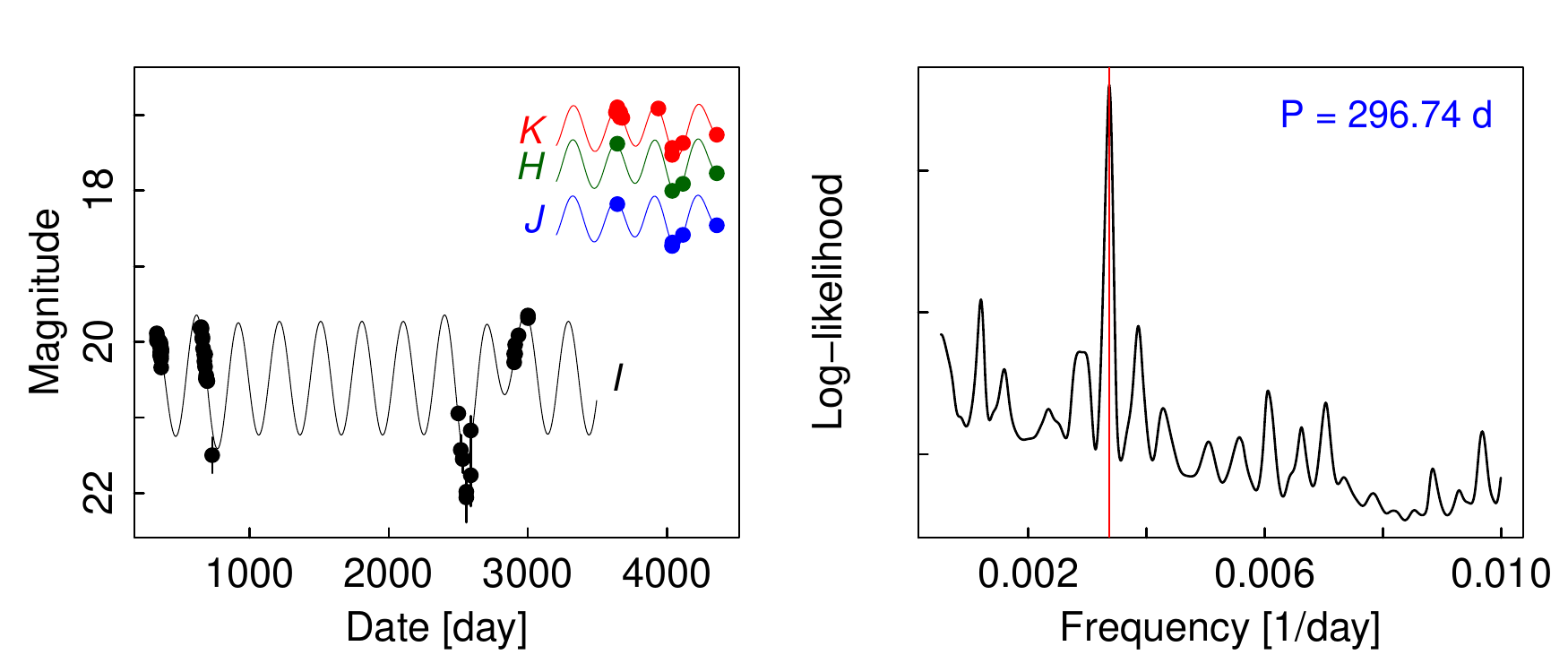}
\caption{Examples of the multiband Gaussian process model fit. {\it Left:} The black, blue, green, and red points indicate the $I$, $J$, $H$, and $K_s$ measurements, respectively. The curves of corresponding colors show the model fit with optimized parameters. {\it Right:} The log-likehood profile of the model fit against trial frequency. The highest peak (red vertical line) is initially adopted as the Mira period. \label{fig_mfit}}
\end{figure*}

\begin{deluxetable}{l@{\extracolsep{5pt}}ccccc}
\tabletypesize{\scriptsize}
\tablecaption{Mira phase lags and amplitude ratios among $I,J,H,K_s$ bands\label{tbl_scale}}
\tablewidth{0pt}
\tablehead{\colhead{} & \multicolumn2c{$a_0$}  & \multicolumn2c{$a_1$} \\ \cline{2-3} \cline{4-5}
& \colhead{O-rich} & \colhead{C-rich} & \colhead{O-rich} & \colhead{C-rich}}
\startdata
 $A_I/A_J$ &   -1.90 &    0.78 &    1.83 &    0.20 \\
 $A_I/A_H$ &   -1.66 &    2.06 &    1.68 &   -0.21 \\
 $A_I/A_K$ &   -1.91 &    4.22 &    1.85 &   -0.91 \\
 $\Delta\Phi_{IJ}/2\pi$ &    0.46 &    0.07 &   -0.14 &   -0.02 \\
 $\Delta\Phi_{IH}/2\pi$ &    0.49 &    0.13 &   -0.15 &   -0.04 \\
 $\Delta\Phi_{IK}/2\pi$ &    0.56 &    0.16 &   -0.17 &   -0.05 \\
\enddata
\tablecomments{These quantities are calculated as $a_0 + a_1 \log(P)$.}
\end{deluxetable}

\vfill\pagebreak\newpage

\subsection{Application to M33 Multiband Data}

We applied the above multiband periodogram procedure to the M33 Mira light curves. We excluded data in a given band obtained with a given telescope if the number of ``epochs'' (measurements separated by more than 5 days) were less than 3. This ensured we could robustly determine zeropoint transformations in a given band across multiple data sources without affecting the periodogram. We note that these rejections were only applied for the period search and not for the rest of the analysis. Figure~\ref{fig_mfit} gives an example of the multiband model fit and the resulting periodogram for one representative Mira candidate. We initially adopted the highest peak in each periodogram as the ``true'' period, but stored the period associated with the second-highest peak for further analysis. The procedure to determine the final choice of period is described in \S~\ref{sec_pcor}.

\begin{figure}
\includegraphics[width=0.49\textwidth]{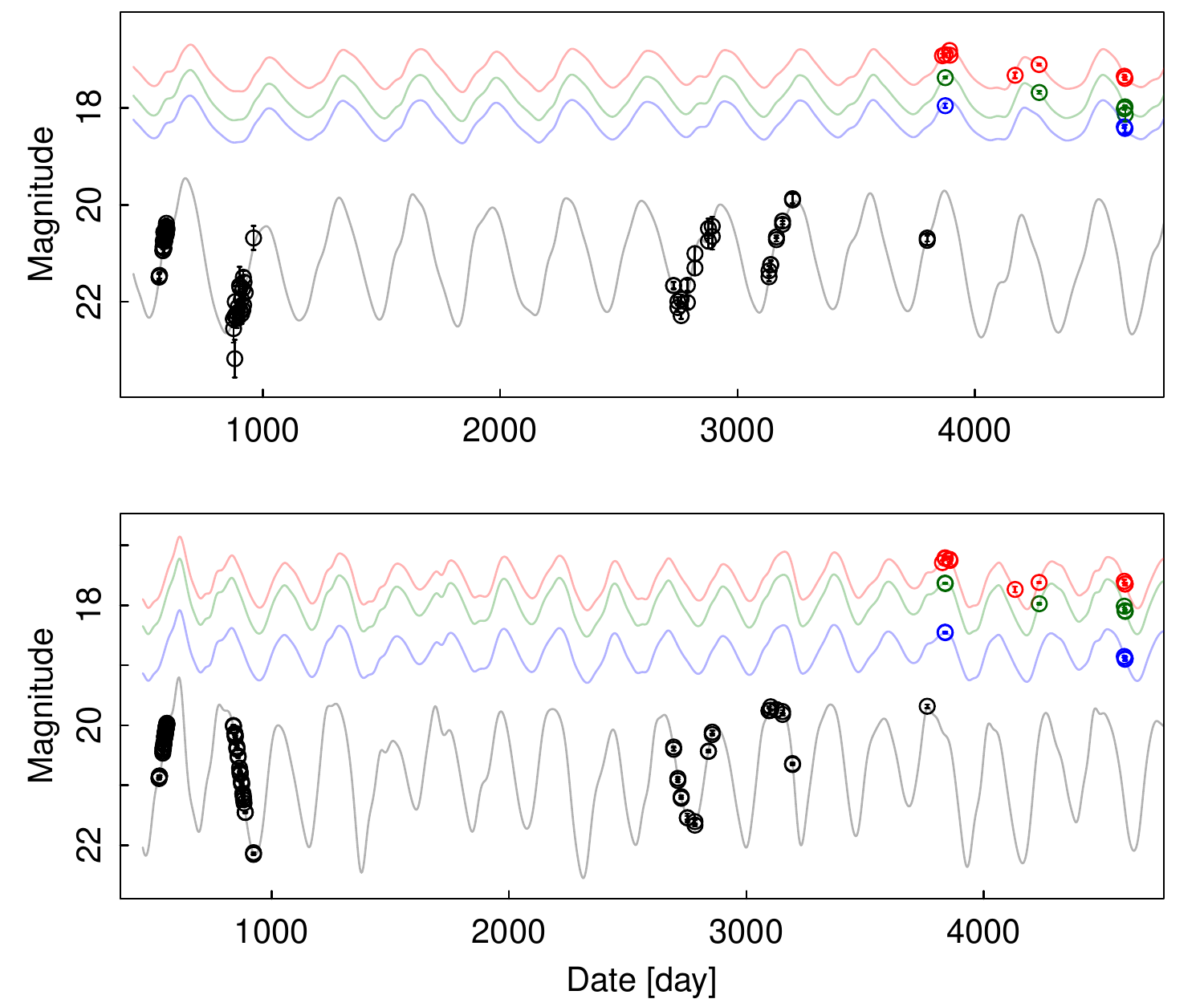}
\caption{Examples of simulated curves with the same sampling. The black, blue, green, and red points indicate simulated measurements in $I$, $J$, $H$, and $K_s$ band, respectively. The continuous curves are derived from the measurements of LMC Miras. \label{fig_simulc}}
\end{figure}

\subsection{Model Accuracy}\label{sec_ma}

To test the accuracy of the model, we simulated $10^4$ multiband Mira light curves (5000 for each subtype) using the LMC $I$-band curves from OGLE-\rom{3}~\citep{Soszynski2009} and the corresponding NIR template curves from~\citet{Yuan2017b}. We drew data points from the LMC Mira curves using the actual sparse sampling patterns of the collected M33 Mira multiband light curves, shifted their magnitudes by +6.27 mag to account for the relative distances between LMC and M33 \citep{Pellerin2011}, and added realistic noise appropriate to each source of photometry. Figure~\ref{fig_simulc} shows two examples of simulated light curves.

\begin{figure*}
\begin{center}
\includegraphics[width=0.8\textwidth]{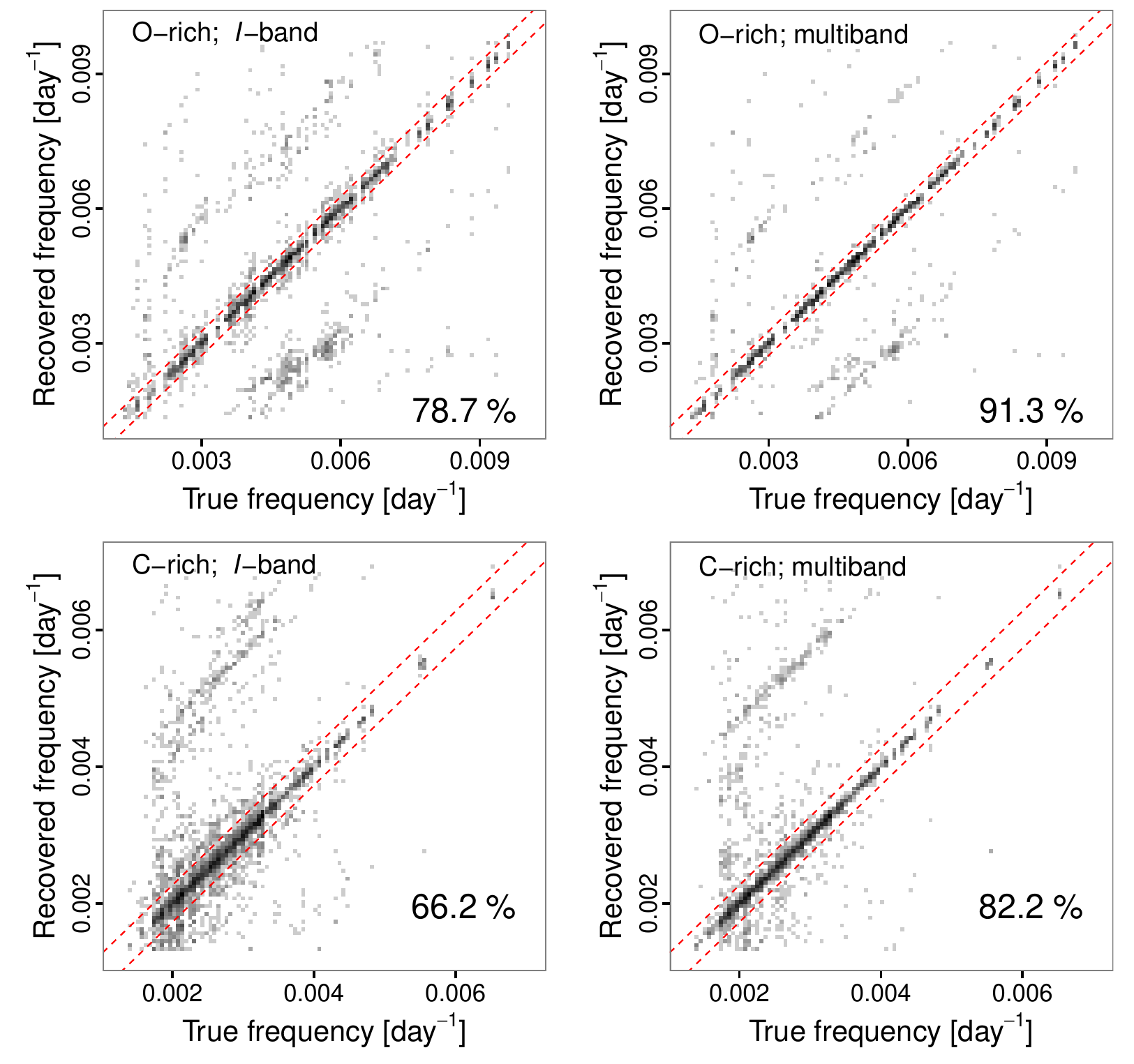}
\end{center}
\caption{Comparison of period recovery rates for the single-band periodogram ($I$, left) and the multi-band formulation ($IJHK_s$, right) using the same simulated Mira light curves. The upper and lower panels show the results for O- and C-rich Mira subtypes, respectively. The percentage of objects within the red bands are labeled in the lower right of each panel. The strong aliasing features parallel to the red bands are due to one-year aliasing frequencies, $\pm $1/365. \label{fig_ff}}
\end{figure*}
 
We applied the multiband periodogram on the simulated data, and compared its performance to the single-band model. We found that the multiband periodogram significantly improves the period recovery rates for both Mira subtypes, as shown in Figure~\ref{fig_ff}. Most of the objects with incorrectly recovered periods fall on two parallel strips in frequency space that correspond to one-year aliasing periods. 

We also used the simulated data to compute the uncertainty of the derived periods. We firstly estimated the relative uncertainties in each Mira candidate using the bootstrap method, computing the period from many subsamples of the measurements. For the simulated data, we computed the same relative uncertainties using identical procedures. Using the differences between recovered and input periods of the simulation, we derived a scale factor to turn relative uncertainties into absolute ones. We finally derived absolute period uncertainties for M33 Mira candidates by applying the scale factor to their relative uncertainties.

\section{Mean Magnitude in NIR}\label{sec_mag}

Given the very limited number of NIR measurements, it is not advisable to solve for mean $JHK_s$ magnitudes using the data-driven model. We used a simpler and more robust method to estimate the mean NIR magnitudes, fitting three sinusoidal curves to the $JHK_s$ measurements. We fixed the amplitude ratios and phase lags among bands, thereby solving for three mean magnitudes, one initial phase, and one absolute amplitude. We then identified C-rich Mira candidates in the M33 Mira sample, and corrected the periods for a subsample of O-rich Mira candidates exhibiting particular PLR residuals.

\subsection{Sinusoidal Fit to NIR Curves}\label{sec_sinmod}

We estimated the mean NIR magnitude by fitting a sinusoidal model to the $JHK_s$ light curves. Unlike the multiband periodogram procedure describe above, we did not exclude any data based on the number of measurements. Instead, we made use of all the NIR data and fit the sinusoidal model to all the $JHK_s$ data simultaneously. Since the number of NIR measurements are generally small, we only included five free parameters in the model, assuming fixed amplitude ratios and phase lags from Table~\ref{tbl_scale}.

For a Mira with period $P\equiv 2\pi/\omega$, the model is
\vspace*{-2pt}
\begin{align*}
J(t) &= a\cdot\cos \omega t + b\cdot\sin\omega t + c \\
H(t) &= a\cdot\cos(\omega t-\Delta\Phi_{JH})\cdot {A_H}/{A_J}\\
     & + b\cdot\sin(\omega t-\Delta\Phi_{JH}) \cdot {A_H}/{A_J} + d \\
K_s(t)& = a\cdot\cos(\omega t-\Delta\Phi_{JK})\cdot {A_K}/{A_J}\\
      & + b\cdot\sin(\omega t -\Delta\Phi_{JK}) \cdot {A_K}/{A_J}  + e
\end{align*}

\begin{figure}
\epsscale{1.1}
\plotone{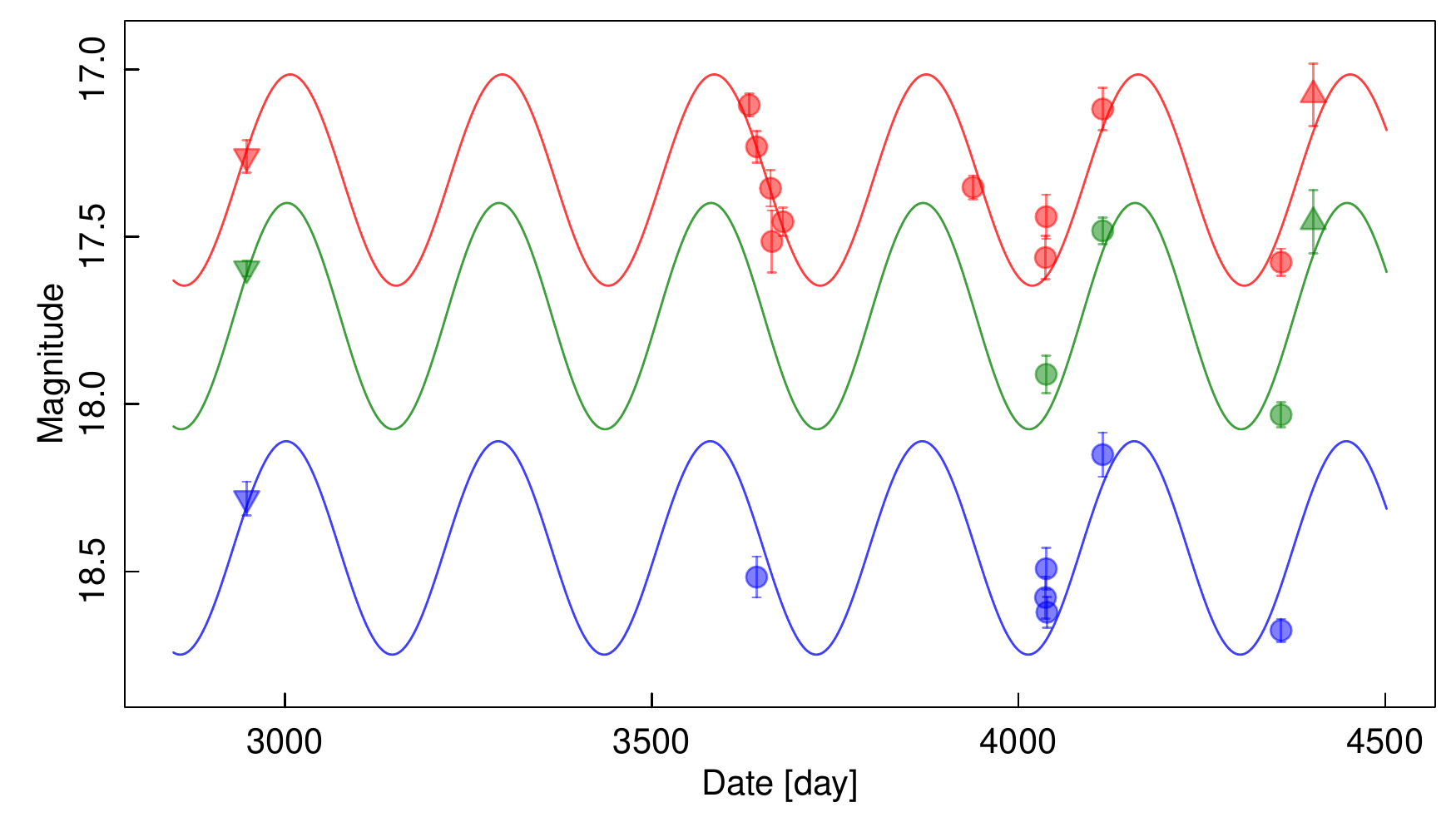}
\caption{Example of the sinusoidal fit to the NIR light curves of Mira candidate w2i2743. The blue, green, and red points indicate the $J$, $H$, and $K_s$ measurements, respectively, while the curves of the corresponding colors shows their best-fit sinusoidal curves. The circles, upward triangles, and downward triangles indicate measurements from UKIRT, KPNO, and Gemini, respectively.\label{fig_fit5}}
\end{figure}

\noindent{where $a$, $b$, $c$, $d$, and $e$ are free parameters and $A_H/A_J$, $A_K/A_J$, $\Phi_{JH}$ and $\Phi_{JK}$ are derived as shown in  Table~\ref{tbl_scale}. We noticed that there are occasionally poor measurements with abnormal magnitudes, and thus fit the model to the NIR light curves using a two-step iterative procedure to exclude those significant outliers. In the first pass, we detected $>3\sigma$ outliers, which were excluded in the second pass to derive the final best-fit parameters. Figure~\ref{fig_fit5} shows an example of the model fit. The mean $JHK_s$ magnitudes were calculated by taking the flux mean of model curves in each band. We computed the uncertainties in magnitude using the same strategy as for the period uncertainty.}

\subsection{Identification of C-rich Miras}\label{sec_cls}

\begin{figure*}
\epsscale{1.1}
\plotone{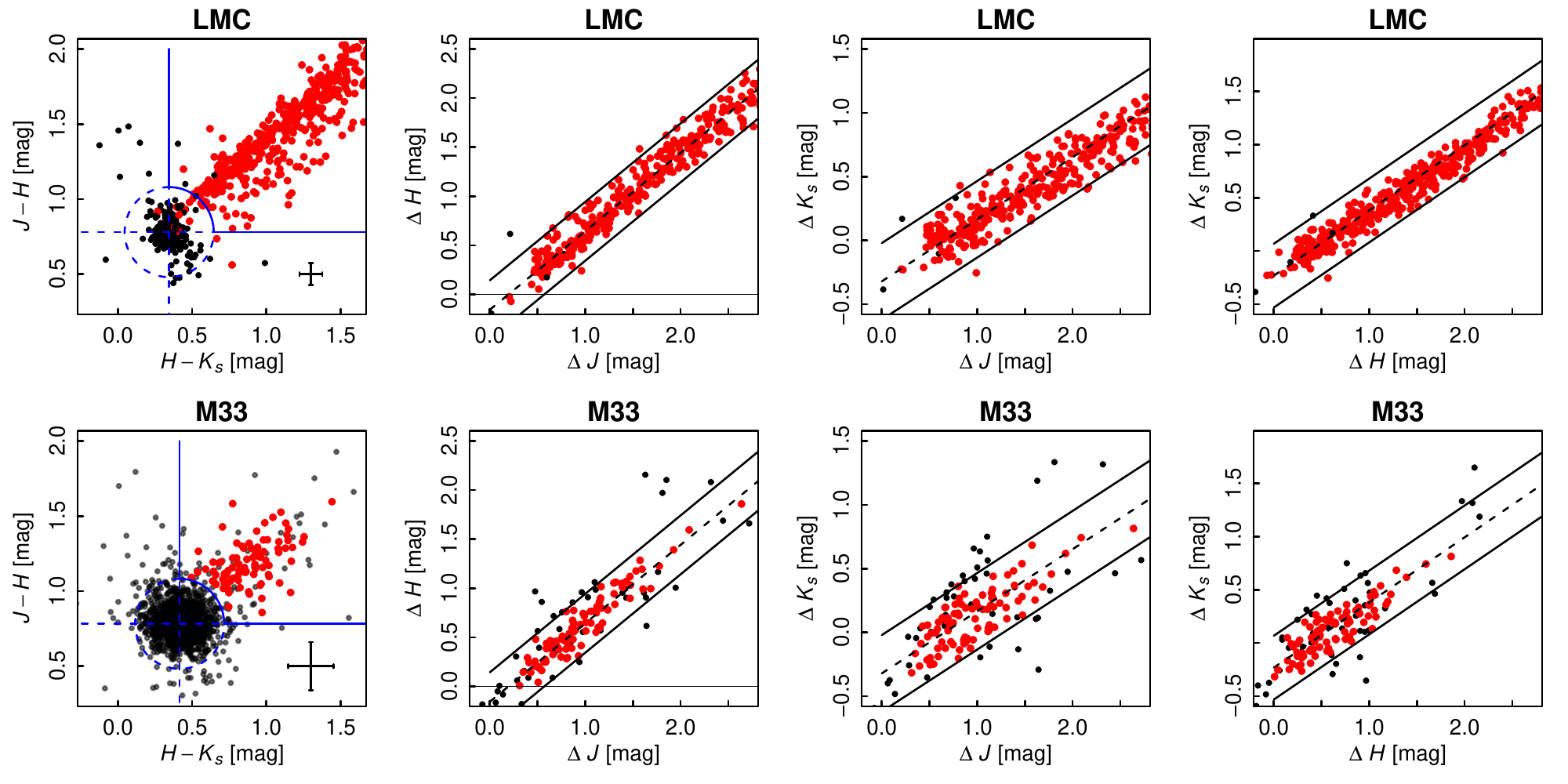}
\caption{Selection of M33 C-rich Mira candidates (lower panels) based on the colors and PLR residuals exhibited by the same type of variables in the LMC (upper panels). For LMC Miras, black and red points indicate O- and C-rich variables. For M33 Miras, the selected C-rich candidates meeting all four selection criteria (solid lines) are indicated by red points. The black points in the rightmost three columns indicate objects that passed the color cut but did not simultaneously meet all three residual relation cuts.\label{fig_oc}}
\end{figure*}

We selected C-rich Mira candidates based on their NIR colors and $JHK_s$ PLR residuals. In~\citet{Yuan2017b} we  demonstrated that C- and O-rich Miras exhibit different $J-H$ and $H-K_s$ color relations and that their PLR residuals are highly correlated across these bands, as shown in the upper panels of Figure~\ref{fig_oc}. We initially selected C-rich candidates in color-color space, requiring that they be located $>0.3$~mag away in the redder direction from the center of the O-rich distribution. This boundary selected 97\% C-rich Miras with $<1$\% contamination. We also required that the PLR residual relations fall within a strip of $\pm 0.3$ mag width for all three combinations of $\Delta J$, $\Delta H$, and $\Delta K_s$. Lastly, we further required $\Delta J > 0$ and $\Delta H > 0$, meaning the objects should be fainter than the mean value for O-rich Miras of the same period.

For each M33 Mira candidate, we computed two sets of NIR colors and $JHK_s$ PLR residuals using the method described in \S~\ref{sec_sinmod} and either C-rich and O-rich $JHK_s$ amplitude ratios. We firstly applied the above cuts using the magnitudes and PLR residuals based on the O-rich relations to select C-rich candidates. We then updated the periods and magnitudes of those selected as C-rich variables using the other set of relations. We noticed that the center of the O-rich distribution in the color-color diagram is slightly different for the LMC sample and M33 sample, and redetermined the center for the M33 sample by iteratively rejecting the one largest outlier until all remaining objects were within a 0.3 mag radius. We found that the M33 sample is centered at $J-H\sim0.78, H-K_s\sim0.42$~mag while the LMC sample is centered at $J-H\sim0.78, H-K_s\sim0.35$~mag. The $\sim$0.07 mag difference in $H-K_s$ color between the two samples is not fully understood, and may be the consequence of contamination by stars other than O-rich Miras in the M33 sample. We used the $JHK_s$ PLR residual relations of LMC C-rich Miras, which did not require modification. We performed the selection of C-rich variables in two passes. We first adopted the M33 distance modulus derived by \citet{Pellerin2011} and determined PLR zeropoints as described in \S~\ref{sec_plr}; in the second pass, we used the updated zeropoints for classification. Using these techniques, we identified 88 C-rich variables out of the 1781 Mira candidates. 

\subsection{Period Correction}\label{sec_pcor}

Based on the simulation described in \S~\ref{sec_ma}, we know that in the case of O-rich Miras there is a $>5\%$~chance that the second-highest peak in the periodogram corresponds to the true period. In such cases, using the primary peak in the periodogram will result in large PLR residuals, while the secondary peak will yield much better agreement.

\begin{figure*}
\includegraphics[width=\textwidth]{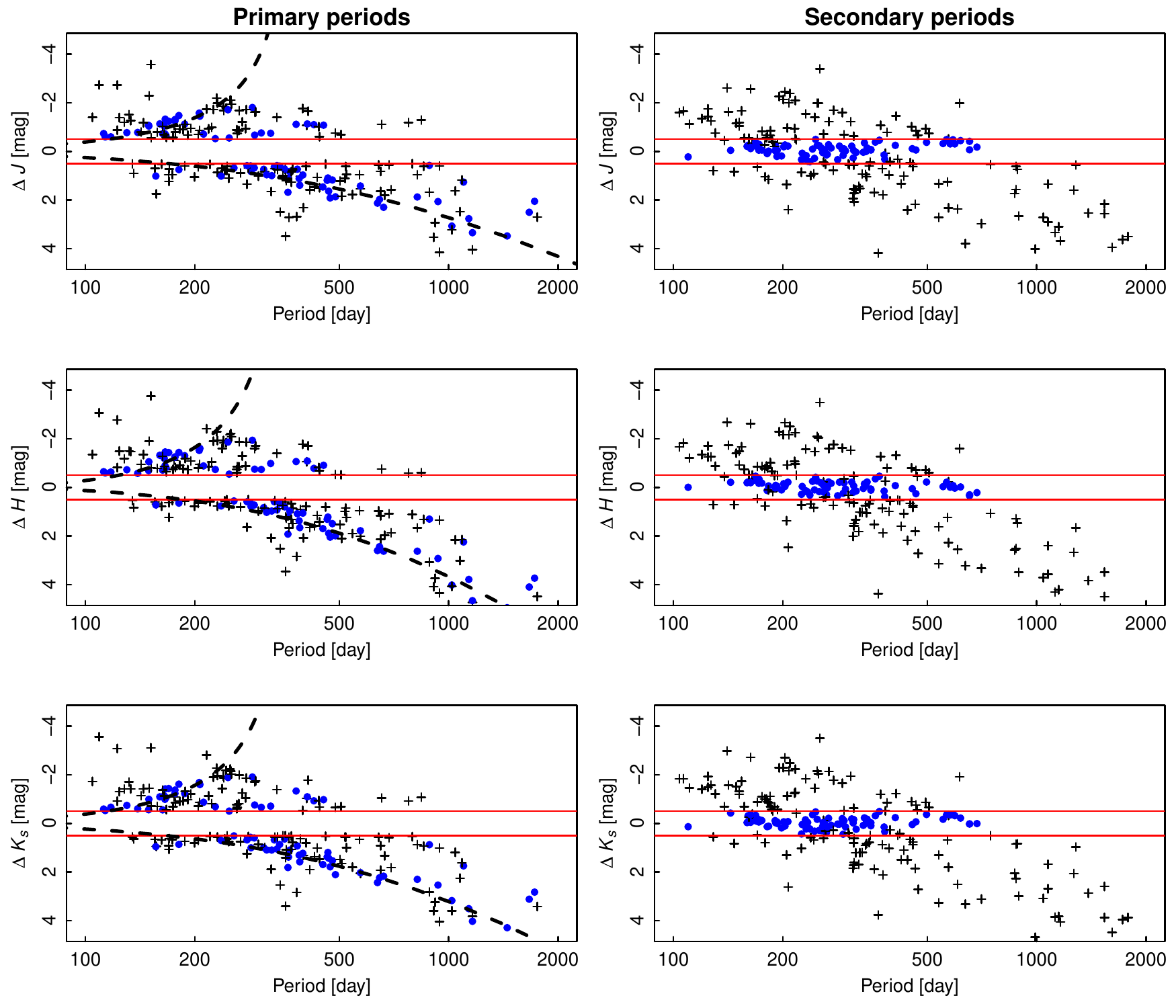}
\caption{PLR residual test for the secondary periods. The left side panels show $JHK_s$ primary-period PLR residuals (from top to bottom) for O-rich Mira candidates with residuals greater than 0.5 mag (indicated by red lines) in all three bands. The right side panels are based on secondary periods. Blue points indicate variables whose secondary periods yield better agreement with the PLRs in all three bands. Black curves indicate one-year aliasing relations. \label{fig_p2}}
\end{figure*}

We therefore computed the PLR residuals of all O-rich Mira candidates (based on the primary peak in the periodogram) and selected outliers beyond $\pm 0.5$ mag in all three bands. We calculated PLR residuals for these objects using the secondary peak of their periodograms, and adopted the alternative period estimate if the residuals were smaller than $\pm 0.5$~mag. This resulted in updated periods for 75 variables, while 135 did not show any significant improvement. Figure~\ref{fig_p2} shows a comparison of PLR residuals using primary and secondary periods. It can be seen that the primary periods of most of the updated variables follow the one-year aliasing relations, which indicates that our correction procedure was well motivated. We performed this procedure in two passes.

\begin{figure}
\epsscale{1.2}
\plotone{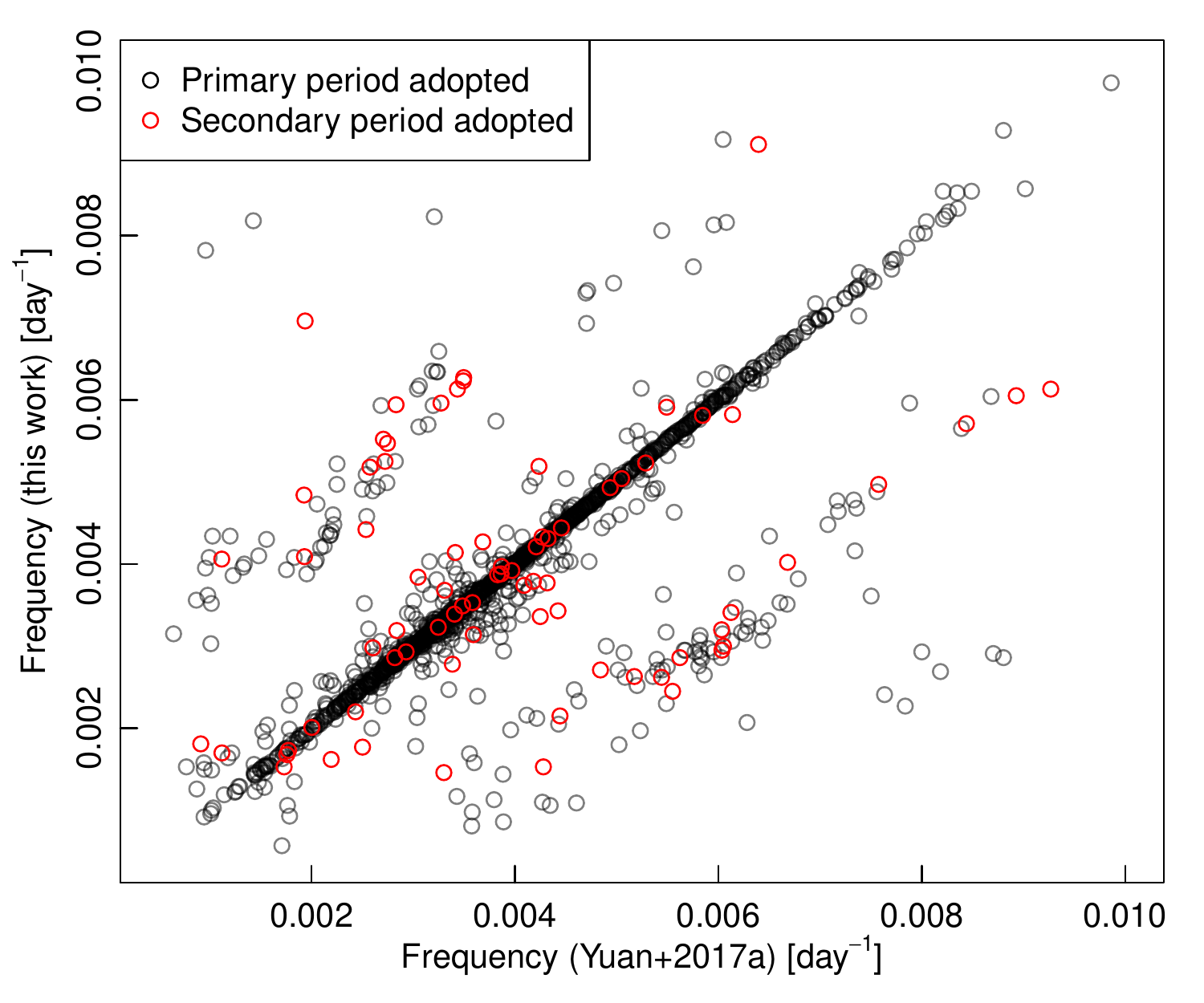}
\caption{Comparison of the periods derived in this work and those obtained by \citet{Yuan2017a}. The black circles indicate objects for which we adopted the primary period, while red circles indicate objects for which we used the secondary periods.\label{fig_cmpp}}
\end{figure}

To summarize, the adopted periods in this study are different from those of~\citet{Yuan2017a} as follows: (1) For objects with adequate NIR time-series measurements, we used the multiband periodogram described in \S3. (2) For O-rich candidates that did not fit the expected Mira PLRs in any of $JHK_s$, we adopted their secondary periods if those fit the PLRs in all three bands. Figure~\ref{fig_cmpp} shows the comparison of the periods used in this study and those derived by \citet{Yuan2017a}.

For the spectroscopically confirmed Mira star [HBS 2006] 40671~\citep{Barsukova2011}, our multiband periodogram gives primary and secondary periods of 426d and 654d, respectively. \citet{Barsukova2011} found a primary period of 665d and secondary periods of 3500d and 406d. This confirmed Mira is one of those objects for which our primary period did not fit the PLRs while the secondary one did; therefore, we adopted the latter as our final choice based on the aforementioned procedure.

\section{Results}\label{sec_result} 

We fit the O-rich Mira PLRs based on LMC variables to the M33 Mira candidates and selected 1265 objects to estimate the distance modulus of M33 and its uncertainty (including systematic errors). 

\subsection{Mira PLRs}\label{sec_plr}

\begin{figure*}
\begin{center}
\includegraphics[width=\textwidth]{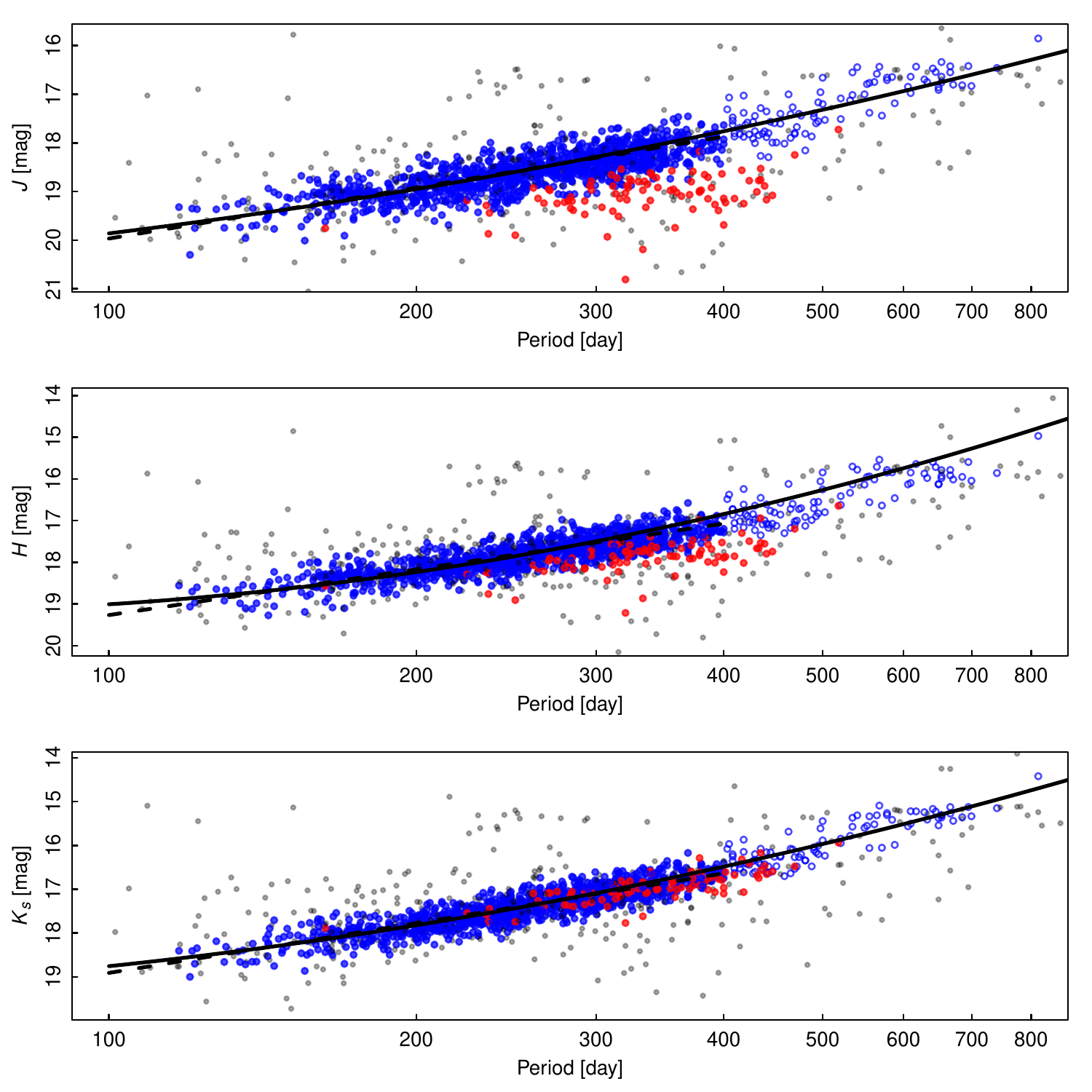}
\end{center}
\caption{M33 Mira PLRs in $J$ (top), $H$ (middle), and $K_s$ (bottom). The blue, red, and gray points indicate O-rich candidates, C-rich candidates, and unclassified candidates, respectively. The dashed and solid lines indicate the PLR fits to the O-rich candidates in first-order and quadratic forms, respectively. The open circles represent variables that were excluded from the fit ($P>400$~d, possible HBB variables).\label{fig_plr}}
\end{figure*}

We fit the quadratic O-rich Mira PLRs from~\citet{Yuan2017b} to the M33 variables of the same subtype. The PLRs are expressed as
\begin{equation*}\label{equ_plr}
M = a_0 + a_1(\log P - 2.3) + a_2(\log P - 2.3)^2
\end{equation*}
where $M$ is the magnitude, $P$ is the period, and $a_{\{012\}}$ are the PLR parameters. We fixed $a_1$ and $a_2$ to the values determined by~\citet{Yuan2017b} and solved for $a_0$. We show the results of the fit in Figure~\ref{fig_plr}. Five types of objects were excluded before or during the fitting process: (1) Objects classified as C-rich Mira candidates ($N=88$); (2) objects with missing magnitudes in any of $JHK_s$ ($N=86$); (3) objects with a problematic fit, indicated by abnormal amplitudes ($A_I>4.5$~mag, $A_J>3$~mag, or $A_J/A_I>1.5$, $N=22$); (4) objects with large period uncertainties ($\sigma_P / P > 0.05$, $N=223$); (5) $>3\sigma$~outliers based on simultaneous iterative clipping across $JHK_s$ ($N=97$). The remaining 1265 objects were classified as O-rich Miras, while the rejected ones were left as unclassified. The 97 Mira candidates that deviated from the PLRs by $> 3 \sigma$ could be misclassified, have incorrect periods, or suffer from very poor measurements (large photometric errors and/or limited sampling); they were excluded from further analysis. We list the properties of all Mira candidates in Table~\ref{tbl_objs}, while the PLR coefficients are given in Table~\ref{tbl_plr}.

Using only O-rich Mira candidates with $P<400$~d, we derived the PLR zeropoints $a_0$ for the linear relations used by~\citet{Yuan2017b}. We did not include the objects with longer periods to avoid any possible contamination by ``hot bottom burning'' variables \citep{Whitelock2003,Marigo2013}. The coefficients of these linear relations are also listed in Table~\ref{tbl_plr}. The scatter of all M33 PLRs is similar for both the linear and quadratic formulations. As described before, the above procedures were performed in a two-step manner, with the second pass using updated results from \S~\ref{sec_cls} and \S~\ref{sec_pcor}.

\subsection{Distance Modulus and Systematic Uncertainty}

We derived the distance modulus of M33 by comparing the zeropoints ($a_0$) of the corresponding LMC and M33 Mira PLRs. We used the offsets of the linear relations for each band, which were corrected for several known sources of bias. We also propagated systematic uncertainties for our estimates.

The difference in computing the ``mean magnitude'' for the LMC and M33 Miras leads to a small but correctable bias. The $JHK_s$ light curves of LMC Miras we fit by \citet{Yuan2017a} using piece-wise templates, and the mean values of the maximum and minimum magnitudes across all segments were used to compute the ``mean magnitude''. This choice was made to avoid significant errors due to template discontinuity. For the M33 measurements, we fit the data with sinusoidal curves and used their flux mean as ``mean magnitude''. We evaluated this bias using the same set of simulated light curves described in \S~\ref{sec_sinmod}, and obtained offsets of 0.034, 0.036, and 0.035 mag for $JHK_s$, respectively.

Another bias comes from the difference in interstellar extinction towards the LMC and M33. For the LMC, we averaged the results of \citet{Haschke2011} based on both red clump stars and RR Lyraes. We used the reddening law from \citet{Fitzpatrick1999} to derive $JHK_s$ extinctions of $A_J=0.06$, $A_H=0.04$, $A_K=0.02$~mag. For M33, we adopted the extinction map from \citet{Schlafly2011}, which gives $A_J=0.03$, $A_H=0.02$, $A_K=0.01$~mag. We corrected the relative distances for this difference in extinction.

The photometric zero point uncertainties are leading factors that contribute to the final error budget. We adopted a conservative 0.02~mag estimate for the internal zeropoint uncertainty of the M33 observations~\citep{Hodgkin2009}. For the LMC measurements, \citet{Macri2015} reports spatially-dependent zeropoint uncertainties. We estimated average values of $\sigma_J\sim 0.03$, $\sigma_H\sim 0.035$, $\sigma_K\sim 0.025$~mag based on their Fig.~4. We added the photometric uncertainties of the two surveys in quadrature and propagated them into the final error budget.

We estimated the bias due to color terms in the photometric calibrations of \citet{Hodgkin2009} for M33 and \citet{Macri2015} for the LMC, which were mostly based on stars bluer than Miras. \citet{Hodgkin2009} reported the WFCAM to 2MASS color terms in their Equations~(4)-(8), while their Figure~10 shows the mean color difference between calibrating stars and O-rich Miras was $\Delta (J-H) \sim 0.3$ and $\Delta (J-K_s) \sim 0.7$ mag. These would bias the distance moduli in $JHK_s$ by 0.02, 0.01, and -0.007~mag, respectively. \citet{Macri2015} reported that the only statistical significant color term was $J=0.018\cdot(J-K)$, and their calibrating stars had a mean $J-K_s=0.99$~mag. This would introduce a bias in the $J$ distance modulus of -0.004~mag.

We did not consider metallicity or differential extinction in this analysis. No observational evidence has been found for a significant metallicity dependence of the NIR PLRs of O-rich Miras \citep{Whitelock1994,Wood1995,Feast1996}. Given the similar abundances of LMC and M33 \citep{Romaniello2008,Bresolin2011}, the overall metallicity effect should be marginal. It is also unlikely that they exhibit significant differential extinction due to circumstellar material, as the intrinsic scatter of the O-rich Mira PLRs is quite small \citep[$\leq 0.12$ mag in $K_s$, see][]{Glass2003,Yuan2017b}.


The aforementioned corrections and associated uncertainties are summarized in Table~\ref{tbl_dist}. We use the LMC distance modulus of $18.493 \pm 0.048$ mag derived by \citet{Pietrzy2013}, correct for all the aforementioned biases, and propagate all uncertainties to arrive at M33 distance moduli of $24.82\pm0.06$, $24.82\pm0.06$, and $24.75\pm0.06$~mag in $JHK_s$, respectively. We average all three values (but maintain the systematic uncertainty in any given band) to arrive at $\mu=24.80\pm 0.06$~mag. This result is somewhat higher, but statistically consistent, with the Cepheid-based distances from the aforementioned studies.

\acknowledgements

W.Y. and L.M.M. acknowledge financial support from NSF grant AST-1211603 and from the Mitchell Institute for Fundamental Physics and Astronomy at Texas A\&M University. This study made use of observations obtained at the Gemini Observatory, which is operated by the Association of Universities for Research in Astronomy, Inc., under a cooperative agreement with the NSF on behalf of the Gemini partnership: the National Science Foundation (United States), the National Research Council (Canada), CONICYT (Chile), Ministerio de Ciencia, Tecnolog\'{\i}a e Innovaci\'{o}n Productiva (Argentina), and Minist\'{e}rio da Ci\'{e}ncia, Tecnologia e Inova\c{c}\~{a}o (Brazil). The authors also acknowledge the Texas A\&M University Brazos HPC cluster that contributed to the research reported here.

\software{DAOPHOT~\citep{Stetson1987}, ALLSTAR~\citep{Stetson1994}, IRAF~\citep{Tody1986, Tody1993}}

\begin{deluxetable*}{lrrrrrrrrrrc}
\tabletypesize{\scriptsize}
\tablecaption{Properties of Mira Candidates\label{tbl_objs}}
\tablewidth{0pt}
\tablehead{
\colhead{ID} & \colhead{R.A.} & \colhead{Dec.} & \colhead{$P$} & \colhead{$\sigma(P)$} &\colhead{$J$} &\colhead{$\sigma(J)$} & \colhead{$H$} & \colhead{$\sigma(H)$} & \colhead{$K_s$} & \colhead{$\sigma(K_s)$} & \colhead{Class$^a$} \\
\colhead{[M33SSSJ]} & & & \multicolumn{2}{c}{(day)} & \multicolumn{2}{c}{(mag)} & \multicolumn{2}{c}{(mag)} & \multicolumn{2}{c}{(mag)} & }
\startdata
01321114+3032588 & 23.04618 & 30.54961 & 325 & 99 & 19.8 & 0.3 & 19.4 & 0.4 & 18.8 & 0.6 & N \\
01321450+3019349 & 23.06024 & 30.32632 & 262 & 2 & 18.5 & 0.1 & 17.89 & 0.05 & 17.53 & 0.05 & O \\
01321654+3025260 & 23.06869 & 30.42384 & 308 & 10 & 18.17 & 0.07 & 17.33 & 0.05 & 16.95 & 0.04 & O \\
01321897+3031226 & 23.07879 & 30.52288 & 256 & 4 & 18.49 & 0.09 & 17.80 & 0.09 & 17.41 & 0.07 & O \\
01322179+3034063 & 23.09052 & 30.56834 & 350 & 3 & 17.9 & 0.2 & 17.3 & 0.2 & 16.8 & 0.2 & O \\
01322351+3030590 & 23.09772 & 30.51630 & 265.2 & 0.7 & 18.5 & 0.1 & 17.78 & 0.09 & 17.33 & 0.07 & O \\
01322586+3033489 & 23.10747 & 30.56352 & 130 & 1 & 19.31 & 0.05 & 18.67 & 0.07 & 18.31 & 0.05 & O \\
01322828+3017589 & 23.11767 & 30.29965 & 337 & 69 & 16.7 & 0.2 & 15.8 & 0.2 & 15.5 & 0.2 & N \\
01322948+3026495 & 23.12265 & 30.44703 & 314 & 1 & 18.12 & 0.03 & 17.33 & 0.04 & 16.88 & 0.04 & O \\
01322979+3034179 & 23.12386 & 30.57156 & 337 & 94 & \multicolumn{2}{c}{$\dots$} & \multicolumn{2}{c}{$\dots$} & 18.2 & 0.5 & N \\
01323105+3031442 & 23.12914 & 30.52887 & 191 & 2 & 19.0 & 0.1 & 18.3 & 0.1 & 17.9 & 0.1 & O \\
01323349+3038395 & 23.13931 & 30.64426 & 500 & 7 & 16.66 & 0.07 & 15.85 & 0.07 & 15.45 & 0.05 & O \\
01323465+3032326 & 23.14412 & 30.54230 & 380 & 20 & 19.0 & 0.5 & 17.7 & 0.4 & 16.6 & 0.3 & C \\
\enddata
\tablecomments{$a$: O for O-rich, C for C-rich, N for not classified.\\(This table is available in its entirety in machine-readable form.)}
\end{deluxetable*}

\begin{deluxetable*}{ccccccccccc}
\tabletypesize{\scriptsize}
\tablecaption{PLR coefficients\label{tbl_plr}}
\tablewidth{0pt}
\tablehead{
\colhead{galaxy} & \colhead{band} & \multicolumn{4}{c}{linear ($P<400$d)}\vline & \multicolumn{5}{c}{quadratic} \\ \cline{3-6} \cline{7-11}
&& \colhead{$a_0$} & \colhead{$a_1$} & \colhead{$\sigma$}  & \colhead{$N$} \vline & \colhead{$a_0$} & \colhead{$a_1$} & \colhead{$a_2$} & \colhead{$\sigma$}  & \colhead{$N$}
}
\startdata
LMC&$  J$&$12.67\pm 0.01$&$-3.48\pm 0.09$& 0.15& 158&$12.70\pm 0.01$&$-3.49\pm 0.09$&$-1.54\pm 0.23$& 0.15& 178\\
M33&$  J$&$18.92\pm 0.01$&$\dots$& 0.25&1169&$18.94\pm 0.01$&$\dots$&$\dots$& 0.25&1265\\
LMC&$  H$&$11.91\pm 0.01$&$-3.64\pm 0.09$& 0.16& 163&$11.96\pm 0.01$&$-3.59\pm 0.10$&$-3.40\pm 0.31$& 0.16& 173\\
M33&$  H$&$18.17\pm 0.01$&$\dots$& 0.24&1169&$18.27\pm 0.01$&$\dots$&$\dots$& 0.26&1265\\
LMC&$K_s$&$11.56\pm 0.01$&$-3.77\pm 0.07$& 0.12& 158&$11.59\pm 0.01$&$-3.77\pm 0.08$&$-2.23\pm 0.20$& 0.12& 176\\
M33&$K_s$&$17.78\pm 0.01$&$\dots$& 0.21&1169&$17.83\pm 0.01$&$\dots$&$\dots$& 0.22&1265\\
\enddata
\end{deluxetable*}

\begin{deluxetable*}{cccccccc}
\tabletypesize{\scriptsize}
\tablecaption{Distance moduli and sources of uncertainty\label{tbl_dist}}
\tablewidth{0pt}
\tablehead{
\colhead{band} & \colhead{$\Delta a_0$}  & \colhead{$\Delta \overline{m}$}  & \colhead{$\Delta A_\lambda$} & \colhead{$\Delta$ct}  & \colhead{$\Delta \mu$}  & \colhead{$\mu_\mathrm{LMC}$}  & \colhead{$\mu$} 
}
\startdata
$J$ & $6.250\pm 0.007$ & $0.034\pm 0.001$ & $0.029\pm 0.008$ & $0.016 \pm 0.036$ & $6.33\pm 0.04$ & $18.493 \pm 0.048$& $24.82\pm 0.06$ \\
$H$ & $6.259\pm 0.007$ & $0.036\pm 0.001$ & $0.018\pm 0.005$ & $0.010 \pm 0.040$ & $6.32\pm 0.04$ & $18.493 \pm 0.048$& $24.82\pm 0.06$ \\
$K_s$ & $6.216\pm 0.006$ & $0.035\pm 0.001$ & $0.012\pm 0.003$ & $-0.007 \pm 0.032$ & $6.26\pm 0.04$ & $18.493 \pm 0.048$ & $24.75\pm 0.06$ \\
\enddata
\tablecomments{$\Delta a_0$: from linear fit (see Table~\ref{tbl_plr}). $\Delta \overline{m}$: correction for calculation of mean magnitude. $\Delta A_\lambda$: correction for differential extinction towards LMC and M33. $\Delta$ct: correction for color terms at mean color of O-rich Miras. $\Delta \mu$: resulting relative distance moduli. $\Delta \mu_\mathrm{LMC}$: Distance modulus to LMC from \citet{Pietrzy2013}. $\mu$: Final distance moduli. All quantities in this table are expressed in magnitudes.}
\end{deluxetable*}


\bibliographystyle{aasjournal}
\bibliography{m33mira}

\appendix

\section{Photometric Transformations}

\begin{figure}
\epsscale{1.0}
\plotone{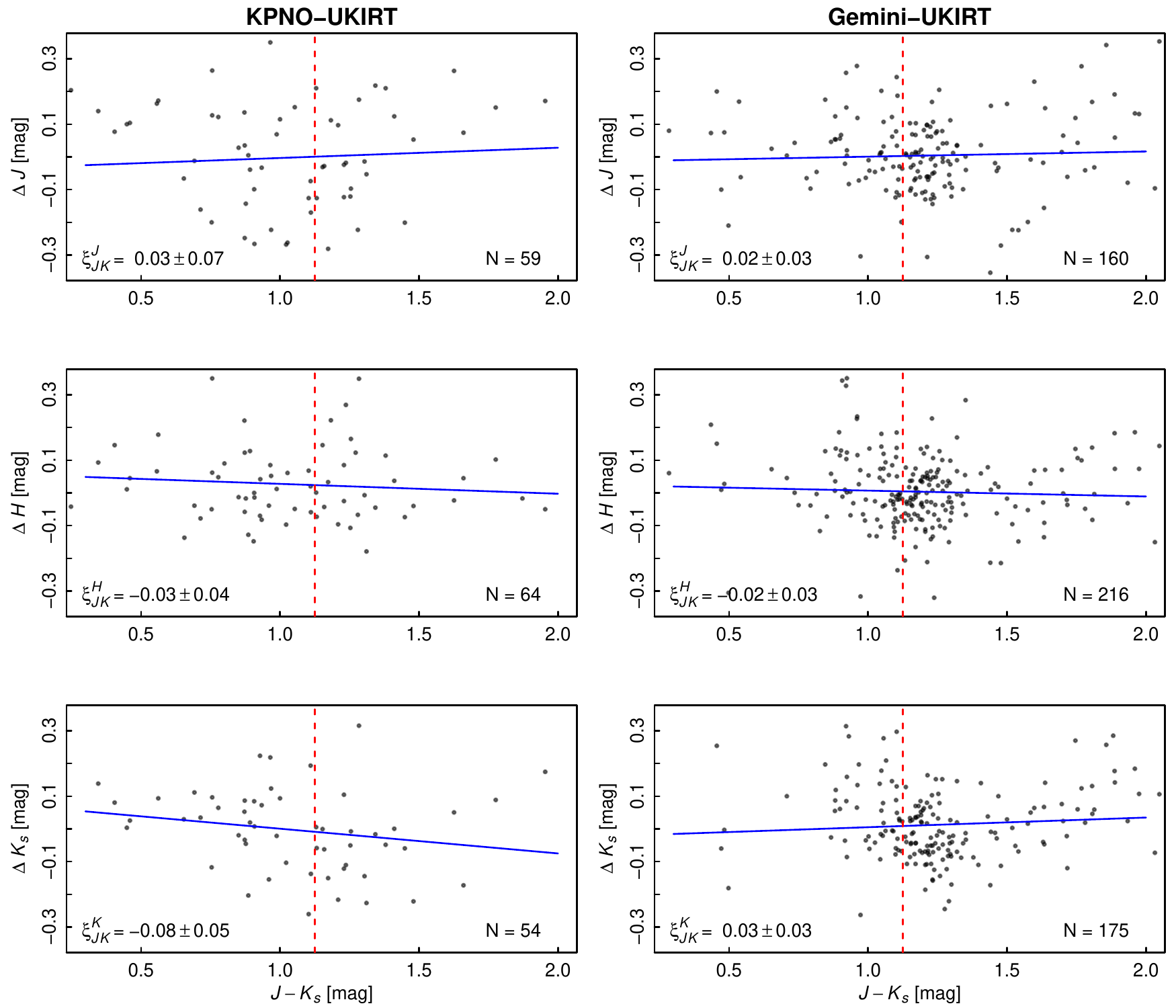}
\caption{Color terms in the photometric calibrations of KPNO/FLAMINGOS (left) and Gemini North/NIRI (right) with respect to UKIRT/WFCAM. The blue solid lines indicate the derived color terms, while the red dashed lines indicate the typical colors of O-rich Miras. The values and uncertainties of the color terms are indicated in the lower-left corner of each panel, while the number of calibrators is given on the lower-right corner.\label{fig_ncolr}}
\end{figure}

We transformed the KPNO/FLAMINGOS and Gemini North/NIRI into the UKIRT/WFCAM system using $\sim 50-200$ bright and isolated stars, depending on the camera and filter. As shown in Figure~\ref{fig_ncolr}, these calibrators span a color range ($0.3 < J-K_s < 2.0$) that brackets the typical color of O-rich Miras. We did not obtain statistically significant color terms for KPNO/FLAMINGOS, given the small number of calibrators and the noisy nature of the photometric measurements. Thus, only a zeropoint correction was applied. In the case of Gemini North/NIRI, the color terms were also small and only marginally more statistically significant, but we included them in the transformation. While C-rich Miras are redder than the calibrating stars and there may exist a systematic error due to the necessary extrapolation, this does not affect our main results as only the O-rich Mira candidates are used for distance determination.

\end{document}